\documentclass[aip,jcp,amsmath,amssymb,reprint,twocolumn,floatfix]{revtex4-2}

\usepackage{url}
\usepackage{graphicx}
\usepackage{dcolumn}
\usepackage{bm}
\usepackage[T1]{fontenc}
\usepackage[version=4]{mhchem}

\begin{document}

\preprint{AIP/123-QED}

\title[Article]{Analytic energy, gradient, and hessian of electrostatic embedding QM/MM based on electrostatic potential fitted atomic charges scaling linearly with the MM subsystem size}

\author{Miquel Huix-Rotllant}
\email{miquel.huixrotllant@univ-amu.fr}
\author{Nicolas Ferr\'e}
\affiliation{Aix-Marseille Univ, CNRS, ICR, Marseille, France.}

\begin{abstract}
Electrostatic potential fitting method (ESPF) is a powerful way of defining atomic charges derived from quantum density matrices fitted to reproduce a quantum mechanical charge distribution in the presence of an external electrostatic potential. These can be used in the Hamiltonian to define a robust and efficient electrostatic embedding QM/MM method. The original formulation of ESPF QM/MM contained two main approximations, namely, the neglect of grid derivatives and the non-conservation of the total QM charge. Here, we present a new ESPF atomic charge operator which solves these drawbacks at virtually no extra computational cost. The new charge operators employ atom-centered grids and conserve the total charge when traced with the density matrix. We present an efficient and easy-to-implement analytic form for the energy, gradient, and hessian that scale linearly with the MM subsystem size. We show that grid derivatives and charge conservation are fundamental to preserve the translational invariance properties of energies and its derivatives and exact conditions to be satisfied by the atomic charge derivatives. As proof of concept, we compute the transition state that leads to the formation of hydrogen peroxide during cryptochrome's reoxidation reaction. Last, we show that the construction of the full QM/MM hessian scales linearly with the MM subsystem size.
\end{abstract}

\maketitle

\section{Introduction}
Quantum mechanics/molecular mechanics (QM/MM) methods are now a routine way to account for the electrostatic interaction between a quantum system and an environment that can be described as an electrostatic scalar external potential.\cite{Warshel1976,Field1990,Lin2007,Liu2014,Senn2009} Three main strategies can account for this interaction: mechanical embedding, electrostatic embedding, and polarizable embedding.\cite{Hai2006,Cao2018} In mechanical embedding, the interacting multipoles are fixed both for the quantum and classical subsystems. In electrostatic embedding, the interacting multipoles are varying for the quantum subsystem and fixed for the classical subsystem. In polarizable embedding, both the quantum and classical multipoles are allowed to vary, usually self-consistently.

The ElectroStatic Potential Fitting (ESPF) method is a general method for the accounting of the QM and MM electrostatic interaction that can be formulated in any of the embedding strategies, from mechanical to polarizable.\cite{Hayashi_00,Ferre02,Melaccio2011} Indeed, ESPF is based on a multipolar expansion of the quantum density of charges. At the lowest order, it can be used to define either classical atomic partial charges that interact with the classical charges (mechanical embedding) \cite{Hu_08} or atomic charge operators\cite{Morita_97,Morita_02,Huang_17} that can be added in the one-electron operators used to solve the QM problem (electrostatic embedding). Being straightforwardly generalized to higher-order multipoles, the ESPF formalism is formally compatible with any type of MM force field and systematically improvable.

The main advantage of the ESPF method over other QM/MM strategies is that the scaling of QM equations is independent of the MM subsystem size.\cite{Schwinn2019} This is possible by defining atomic multipole operators in the QM subsystem based on the fitting of electrostatic integrals computed on a QM-centered regular grid. This implies that the QM subsystem only ``knows'' about the MM environment by means of its electrostatic potential. Consequently, the QM equations avoid the scaling with the MM atom number for the energy and its derivatives. Indeed, the derivative of the ESPF operators due to a perturbation of the MM environment only has a contribution from the electron density polarization. We recently showed a strategy to efficiently obtain this contribution\cite{Schwinn2019}, opening the way to the first complete and efficient calculation of the full QM/MM energy, gradient and hessian in a QM/MM electrostatic embedding framework.\cite{Schwinn2020,Schwinn2020b}

The original ESPF formulation in Ref.\ \citenum{Ferre02} contained two main approximations: (i) the charge of the QM subsystem was not exactly conserved, inherent to the finite size of the grid on which electrostatic integrals are calculated, and (ii) the grid derivative contributions were missing for the energy derivatives. The computation of grid-based atomic charges offer not only an obvious computational advantage for the calculation of electrostatic integrals, but also the fact that atomic charges only depend on the coordinates of the QM subsystem. The latter property has been used several times to derive charges fitted to the electrostatic potential calculated on a grid due to the QM distribution of charges.\cite{Singh1984,Besler90,Sigfridsson1998,Levy_98,Dognon_00,Tu_01,Chen_10} However, depending how the grid is constructed and whether grid derivatives (with respect to QM atom coordinates) are included, it can have severe consequences for sum-rules of atomic charges and the translational and rotational invariance of atomic charges.\cite{Breneman1990} In electrostatic embedding QM/MM, this symmetry breaking is introduced in the Hamiltonian and the resulting energy derivatives, similar to what it was observed in density-functional theory (DFT) exchange-correlation functional integration on numerical grids.\cite{Johnson1994,Malagoli2003}

Here, we present a new ESPF formulation based on atom-centered Lebedev grids with no approximations to a given order of multipolar expansion. To the lowest order of this expansion, we define a new atomic charge operator that conserve the total charge of the QM subsystem without employing Lagrange multipliers, and includes grid derivative contributions for its derivatives. In all formulas for constructing the gradient and hessian, we avoid the derivative of the electrostatic kernel pseudoinverse. This boosts the efficiency of the ESPF computations, allowing routine energy, gradient and hessian calculations at the electrostatic embedding QM/MM level.

\section{Methodology}
 
\subsection{Electrostatic Potential fitting method}

The main equations of the electrostatic embedding ESPF method were presented elsewhere.\cite{Ferre02,Melaccio2011,Schwinn2019,Schwinn2020,Schwinn2020b} For completeness, we summarize the main equations for the energy expression, first and second derivatives. In the ESPF method, the electrostatic kernel (${\bf T}$) plays a central role. To the lowest order, it is defined as a Coulomb interaction kernel between a quantum atom $B$ and a grid point $k$ around atom B,
\begin{equation}
 T_{k;B}=\frac{1}{|{\bf r}_k-{\bf r}_B|}\label{eq:t} \, .
\end{equation}
This allows a simple definition of a charge operator based on the on-grid electrostatic integrals to the same order $V_{k;\mu\nu}=\langle\chi_\mu|\left|r-r_k\right|^{-1}|\chi_\nu\rangle$, in which $\chi_\mu$ is an atomic orbital. The ESPF atomic charge operator is thus obtained by solving the linear set of equations,\cite{Ferre02}
\begin{equation}
 \sum_k^{\mathrm{N}_\mathrm{grid}}{\sum_A^{\mathrm{N}_\mathrm{QM}}{T_{B;k} w_{k}T_{k;A}{\hat Q}_{A,\mu\nu}}} = \sum_k^{\mathrm{N}_\mathrm{grid}}{T_{B;k}w_{k}V_{k;\mu\nu}} \label{eq:wtqv} \, .
\end{equation}
Here ${\bf w}$ are the weights of the grid, $\mathrm{N}_\mathrm{grid}$ are the number of grid points and $\mathrm{N}_\mathrm{QM}$ are the number of atoms treated at the quantum level of theory. Here, ${\hat Q}$ correspond to operator leading to the electronic net population (for simplicity, hereafter it will be referred as atomic charge operator). The electrostatic kernel forms a rectangular matrix, that can be inverted using the Moore-Penrose inverse technique,\cite{Penrose1955} which allows us to write the charge operator as,\cite{Ferre02}
\begin{eqnarray}
 {\hat Q}_{A,\mu\nu} &=& \sum_k^{\mathrm{N}_\mathrm{grid}}{\sum_B^{\mathrm{N}_\mathrm{QM}}{\left(T_{A;k} w_{k}T_{k;B}\right)^{-1}\sum_{k'}^{\mathrm{N}_\mathrm{grid}}{T_{B;k'} w_{k'}V_{k';\mu\nu}}}} \nonumber \\
 &:=&\sum_k^{\mathrm{N}_\mathrm{grid}}{[T_w^+]_{A;k}V^{\vphantom{\dagger}}_{k;\mu\nu}}
 \label{eq:qop} \, .
\end{eqnarray}
Hereafter, we use the common notation for pseudoinverse matrices ${\bf T}_{\bf w}^+=({\bf T}^\dagger{\bf w}{\bf T})^{-1}{\bf T}^\dagger{\bf w}$. Using this definition for the charge operator, we can easily define electronic net populations centred on the quantum atoms $A$, by tracing the charge operator with the density matrix ({\bf P}),
\begin{equation}
 Q_A=\sum_{\mu\nu}^{\mathrm{N}_\mathrm{AO}}{P_{\mu\nu}{\hat Q}_{A;\mu\nu}}=\mathrm{Tr}_\mathrm{AO}\left[{\bf P}{\hat {\bf Q}}_A\right] \, .
 \label{eq:qespf}
\end{equation}
These charges can interact with the classically defined external potential $\phi_A$, giving a total energy contribution
\begin{equation}
 \Delta E=\sum_A^{\mathrm{N}_\mathrm{QM}}{\left(Z_A-Q_A\right)\phi_A}=\mathrm{Tr}_\mathrm{QM}[{\bf q}\pmb{\phi}] \, ,
 \label{eq:e0espf}
\end{equation}
in which $Z_A$ is the nuclear charge and $q_A=Z_A-Q_A$ is the atomic partial charge of atom A. This energy can be used to define a mechanical embedding approach directly. Alternatively, for an electrostatic embedding scheme in which the density matrix is polarized by the electrostatic environment, one can define an ESPF operator
\begin{equation}
 h_{\mu\nu}=\sum_A^{\mathrm{N}_\mathrm{QM}}{\phi_A{\hat Q}_{A,\mu\nu}} \, ,
 \label{eq:espfop}
\end{equation}
that can be added in the QM Fock operator.

\paragraph{Energy gradient:}
The first derivative of the ESPF interaction energy\cite{Ferre02,Melaccio2011,Schwinn2020} given in eq.\ \ref{eq:e0espf} with respect to an atom $C$ can be easily written as
\begin{equation}
 \Delta E^{x_C}=-\mathrm{Tr}_\mathrm{QM}[{\bf Q}^{x_C}\pmb{\phi}]+\mathrm{Tr}_\mathrm{QM}[{\bf q}\pmb{\phi}^{x_C}] \, .
 \label{eq:e1espf}
\end{equation}
Hereafter we use the derivative notation $\Delta E^{x_C}\equiv\partial\Delta E/\partial x_C$. Equation \ref{eq:e1espf} has a simple physical interpretation. The first term involves the charge fluctuations due to the fact that the quantum density is polarizable. The second term correspond to the fluctuations of the external field at fixed atomic charges, and has a trivial expression since the external field has a classical analytic form. Here, we concentrate on the first term, involving the derivative of the ESPF charge operator. Deriving eq.\ \ref{eq:qespf}, we obtain
\begin{equation}
 {\bf Q}_A^{x_C}=\mathrm{Tr}_\mathrm{AO}[{\bf P}^{x_C}{\hat {\bf Q}_A}]+\mathrm{Tr}_\mathrm{AO}[{\bf P}{{\hat {\bf Q}}}_A^{x_C}] \, .
 \label{eq:q1}
\end{equation}
The first term corresponds to the dipole induced by the fluctuations of the density matrix, while the second corresponds to the induced dipole operator at fixed density. The first term requires the solution of the coupled perturbed self-consistent field equations (CPSCF). In general, this has to be solved for all MM and QM atomic motions, although this can be avoided for the computation of gradients.\cite{Schwinn2019} The second term involves the first-order derivative of the ESPF charge operator, which is given
\begin{eqnarray}
{\hat {\bf Q}}^{x_C}&=&{\bf T}^+_w\left[{\bf V}^{x_C}-{\bf T}^{x_C}{\bf Q}\right] \nonumber \\
&+&\left({\bf T}^\dagger{\bf w}{\bf T}\right)^{-1}\left[{\bf T}^\dagger{\bf w}\right]^{x_C}\left[{\bf V}-{\bf T}{\bf Q}\right] \, .
\label{eq:q1op}
\end{eqnarray}
Thus, the derivative of the charge operator requires the derivative of the on-grid electrostatic integrals and the electrostatic kernel. On the one hand, the integral derivative have an easy expression
\begin{eqnarray}
 V^{x_C}_{k;\mu\nu}=\langle\chi_\mu^{x_C}|\frac{1}{\left|{\bf r}-{\bf r}_{k}\right|}|\chi_\nu\rangle+\langle\chi_\mu|\frac{1}{\left|{\bf r}-{\bf r}_{k}\right|}|\chi^{x_C}_\nu\rangle \, ,
 \label{eq:v1}
\end{eqnarray}
in which the derivative of the electrostatic kernel have been neglected. Here, $\chi_\mu$ symbolizes an atomic orbital, and the bra-ket notation for integral should be interpreted as $\langle\chi_\mu|{\left|{\bf r}-{\bf r}_{k}\right|}^{-1}|\chi_\nu\rangle=\int{d{\bf r}\chi^*_\mu({\bf r})\left|{\bf r}-{\bf r}_{k}\right|^{-1}\chi_\nu({\bf r})}$. On the other hand, the derivative of the electrostatic kernel is given by
\begin{equation}
 T_{k;B}^{x_C}=\delta_{BC}\frac{x_k-x_B}{\left | {\bf r}_k - {\bf r}_B \right |^3} \, .
 \label{eq:t1}
\end{equation}
Here $\delta$ symbolizes a Kronecker's delta. Again, the grid derivatives have been neglected.

\paragraph{Energy hessian:} 

The second derivative of the ESPF interaction energy \cite{Schwinn2020,Schwinn2020b} given in eq.\ \ref{eq:e0espf} with respect to an atom C and an atom D can be written as
\begin{eqnarray}
 \Delta E^{x_Cy_D}&=&-\mathrm{Tr}_\mathrm{QM}[{\bf Q}^{x_Cy_D}\pmb{\phi}]+\mathrm{Tr}_\mathrm{QM}[{\bf q}\pmb{\phi}^{x_Cy_D}] \nonumber \\
 &+&\mathrm{Tr}_\mathrm{QM}[{\bf Q}^{y_D}\pmb{\phi}^{x_C}]-\mathrm{Tr}_\mathrm{QM}[{\bf Q}^{x_C}\pmb{\phi}^{y_D}] \, .
 \label{eq:e2espf}
\end{eqnarray}
The first term corresponds to an induced quadrupole at fixed external potential, the second term corresponds to the second order variations of the external field at fixed atomic charges, and the last two terms correspond to the induced dipoles interacting with a first-order variation of the external potential. The second term has an easy analytic expression, while the third and fourth terms are constructed with information from the energy gradient. Therefore, here we concentrate on the expression of the induced quadrupole, which can be expressed as
\begin{eqnarray}
{\bf Q}_A^{x_Cy_D}&=&\mathrm{Tr}_\mathrm{AO}[{\bf P}^{x_Cy_D}{\hat {\bf Q}}_A]+\mathrm{Tr}_\mathrm{AO}[{\bf P}{\hat {\bf Q}}_A^{x_Cy_D}] \nonumber \\
&+&\mathrm{Tr}_\mathrm{AO}[{\bf P}^{y_D}{\hat {\bf Q}}_A^{x_C}]+\mathrm{Tr}_\mathrm{AO}[{\bf P}^{x_C}{\hat {\bf Q}}_A^{y_D}] \, .
\label{eq:q2}
\end{eqnarray}
For constructing it, the first and second derivatives of the density matrix and the charge operator are required. Similar to the energy gradient, one can avoid the computation of the second derivative of the density matrix, but the first derivative density matrix is required, for which a set of CPSCF equations have to be solved to compute it. For the ESPF hamiltonian, this can be efficiently done for the QM and MM perturbations, avoiding the solution of CPSCF for each MM atom (for further details see Ref. \citenum{Schwinn2019}). The expression for the second derivative of the atomic charge operator is given by
\begin{eqnarray}
{\hat {\bf Q}}^{x_Cy_D}&=&{\bf T}^+_w\left[{\bf V}^{x_Cy_D}-{\bf T}^{x_C}{\bf Q}^{y_D}-{\bf T}^{y_D}{\bf Q}^{x_C}-{\bf T}^{x_Cy_D}{\bf Q}\right] \nonumber \\
&+&\left({\bf T}^\dagger{\bf w}{\bf T}\right)^{-1}\left[{\bf T}^\dagger{\bf w}\right]^{x_Cy_D}\left[{\bf V}-{\bf T}{\bf Q}\right] \nonumber \\
&+&\left({\bf T}^\dagger{\bf w}{\bf T}\right)^{-1}\left[{\bf T}^\dagger{\bf w}\right]^{x_C}\left[{\bf V}-{\bf T}{\bf Q}\right]^{y_D} \nonumber \\
&+&\left({\bf T}^\dagger{\bf w}{\bf T}\right)^{-1}\left[{\bf T}^\dagger{\bf w}\right]^{y_D}\left[{\bf V}-{\bf T}{\bf Q}\right]^{x_C} \, ,
\label{eq:qop2}
\end{eqnarray}
Similar to the gradient, the second derivative of the atomic charge operator requires the information of the first-derivative of the operator in addition to the second-derivative of the electrostatic kernel pseudoinverse and the on-grid integrals. On the one hand, the expression for the second-derivative integrals is given by
\begin{eqnarray}
 V^{x_Cy_D}_{k;\mu\nu}&&=\langle\chi_\mu^{x_Cy_D}|\frac{1}{\left|{\bf r}-{\bf r}_{k}\right|}|\chi_\nu\rangle+\langle\chi_\mu|\frac{1}{\left|{\bf r}-{\bf r}_{k}\right|}|\chi^{x_Cy_D}_\nu\rangle
 \nonumber \\
 &+& \langle\chi_\mu^{x_C}|\frac{1}{\left|{\bf r}-{\bf r}_{k}\right|}|\chi^{y_D}_\nu\rangle+\langle\chi^{y_D}_\mu|\frac{1}{\left|{\bf r}-{\bf r}_{k}\right|}|\chi^{x_C}_\nu\rangle\, ,
 \label{eq:v2}
\end{eqnarray}
in which the grid derivatives have been neglected. On the other hand, the second derivative of the electrostatic kernel, which is given by
\begin{equation}
 T_{k;B}^{x_Cy_D}=\delta_{CB}\delta_{DB}\frac{(x_k-x_B)(y_k-y_B)-\delta_{xy}\left | {\bf r}_k - {\bf r}_B \right |^2}{\left | {\bf r}_k - {\bf r}_B \right |^5} \, .\label{eq:t2}
\end{equation}
Again, the grid derivatives have been neglected.

\subsection{ESPF based on atom-centered grids}

Here, we reformulate the original ESPF method by using atom-centered grids.\cite{Holden2013} The expressions for the interaction energy (eq.\ \ref{eq:e0espf}) and its first (eqs. \ref{eq:e1espf} and \ref{eq:q1}) and second derivatives (eqs. \ref{eq:e2espf} and \ref{eq:q2}) remain valid. Therefore, here we concentrate on the expression of the first and second derivatives of the ESPF charge operator and the extra contributions that arise from the grid derivatives. First of all, we define a grid centred on the QM atoms. We can redefine the previous grid points $k$ by a new grid definition of points of a Lebedev sphere centered on quantum atoms $A$,
\begin{equation}
 {\bf r}_k = {\bf r}_A + {\bf r}_l := {\bf r}_{A,l} \, .
 \label{eq:lebgrid}
\end{equation}
In this section, we consider all grid points contribute equally, so that the grid weights are considered all equal to 1. The advantage of using an atom-centred grid is that we can compute the grid derivatives with respect to a quantum atom \textit{B} is simply given by ${\bf r}^{x_B}_{A,l}=\delta_{AB}{\bf e}_x$, in which ${\bf e}_x$ is the x-axis standard unit vector. Accordingly, all grid derivatives can be expressed as QM atom position derivatives. In this atom-centered grid, the electrostatic kernel (eq. \ref{eq:t}) can be re-expressed as
\begin{equation}
 T_{A,l;B}=\frac{1}{|{\bf r}_{A,l}-{\bf r}_B|} \, .
\end{equation}
Using this grid, one can re-write the atomic charge operator elements (eq.\ \ref{eq:qespf}) as
\begin{eqnarray}
 {\hat Q}_{A,\mu\nu}&=&\sum_{B}^{\mathrm{N}_\mathrm{QM}}{\sum_{l}^{\mathrm{N}_\mathrm{B,l}}{\left[T_w^+\right]_{A;B,l}V_{B,l;\mu\nu}}} \nonumber \\
 &:=& \sum_{B,l}{\left[T_w^+\right]_{A;B,l}V_{B,l;\mu\nu}} \, .
\end{eqnarray}
In the second equation, we defined the short-hand sum over $B$ and $l$ for the grid. Indeed, these two sums do not commute, since the number of Lebedev grid points per atom ($\mathrm{N}_\mathrm{B;l}$) is different for each atom.

\paragraph{Charge operator gradient:}
The first derivative of the electrostatic kernels with respect to an atom $C$ are $T^{x_C}_{A,l;C}=-T^{x_C}_{C,l;A}$ are non-zero for $A\ne C$, and zero otherwise. It is worth noting that the ``local'' electrostatic kernel of atom \textit{A} is independent of its position $T_{A,l;A}=|{\bf r}_l|^{-1}$ and $T_{A,l;A}^{x_C}=0$ for $\forall$ C. The first derivative kernel in eq.\ \ref{eq:t1} can now be written as
\begin{equation}
 T_{A,l;B}^{x_C} = \left ( \delta_{BC} - \delta_{AC} \right )\frac{x_{A,l}-x_B}{\left | {\bf r}_{A,l} - {\bf r}_B \right |^3} \, ,
 \label{eq:t1grid}
\end{equation}
including the grid derivatives, which introduces the second delta function in the equation. The first derivatives of the on-grid electrostatic integrals defined on atom-centred grids is given by
\begin{eqnarray}
 V^{x_C}_{B,l;\mu\nu}&=&\langle\chi_\mu^{x_C}|\frac{1}{\left|{\bf r}-{\bf r}_{B,l}\right|}|\chi_\nu\rangle+\langle\chi_\mu|\frac{1}{\left|{\bf r}-{\bf r}_{B,l}\right|}|\chi^{x_C}_\nu\rangle \nonumber \\ &+&\delta_{BC}\langle\chi_\mu|\frac{\left(x-x_{B,l}\right)}{\left|{\bf r}-{\bf r}_{B,l}\right|^3}|\chi_\nu\rangle \, .
 \label{eq:v1grid}
\end{eqnarray}
The last terms both in eq.\ \ref{eq:t1grid} and \ref{eq:v1grid} arise from the grid derivatives, and can be interpreted as the on-site induced dipole moments due to the variations of atom $C$. Using these definitions in eq.\ \ref{eq:q1op}, one obtains the full expression for the gradient of the charge operator including grid derivatives (see eq. S1 in the Supporting Information). Equivalent expressions for the derivative of the charges were obtained by \citeauthor{Holden2019} using CHELPG charges.\cite{HoldenPhD,Holden2019} However, our formula avoid the pseudoderivative of the inverse of the electrostatic kernel, which makes the derivative for the charges and the implementation simpler and lessens the computational cost for calculating charge derivatives. While the difference is not large for the gradient, it becomes important for the hessian.

\begin{widetext}
\paragraph{Charge operator hessian:}
The second derivative of the electrostatic kernel previously defined in eq.\ \ref{eq:t2} can now be written as
\begin{equation}
 T_{A,l;B}^{x_Cy_D} = 
 \left ( \delta_{BC} - \delta_{AC} \right )\left ( \delta_{BD} - \delta_{AD} \right )\frac{\left(x_{A,l}-x_B\right)\left(y_{A,l}-y_B\right)-\delta_{xy}\left| {\bf r}_{A,l} - {\bf r}_B \right|^2}{\left | {\bf r}_{A,l} - {\bf r}_B \right |^5} \, ,\nonumber
\end{equation}
thus including three extra terms due to the grid derivatives. The second derivatives of the electrostatic integrals on a grid are given by
\begin{eqnarray}
 V^{x_Cy_D}_{B,l;\mu\nu}&=&\langle\chi_\mu^{x_Cy_D}|\frac{1}{|{\bf r}-{\bf r}_{B,l}|}|\chi_\nu\rangle+\langle\chi_\mu|\frac{1}{|{\bf r}-{\bf r}_{B,l}|}|\chi^{x_Cy_D}_\nu\rangle+\langle\chi_\mu^{x_C}|\frac{1}{|{\bf r}-{\bf r}_{B,l}|}|\chi^{y_D}_\nu\rangle+\langle\chi^{y_D}_\mu|\frac{1}{|{\bf r}-{\bf r}_{B,l}|}|\chi^{x_C}_\nu\rangle \nonumber \\
 &+&\delta_{BD}\left[\langle\chi_\mu^{x_C}|\frac{\left(y-y_{B,l}\right)}{\left|{\bf r}-{\bf r}_{B,l}\right|^3}|\chi_\nu\rangle+\langle\chi_\mu|\frac{\left(y-y_{B,l}\right)}{\left|{\bf r}-{\bf r}_{B,l}\right|^3}|\chi^{x_C}_\nu\rangle
 \right]+\delta_{BC}\left[\langle\chi_\mu^{y_D}|\frac{\left(x-x_{B,l}\right)}{\left|{\bf r}-{\bf r}_{B,l}\right|^3}|\chi_\nu\rangle+\langle\chi_\mu|\frac{\left(x-x_{B,l}\right)}{\left|{\bf r}-{\bf r}_{B,l}\right|^3}|\chi^{y_D}_\nu\rangle\right] \nonumber \\
 &+&\delta_{BC}\delta_{BD}\langle\chi_\mu|\frac{(x-x_{B,l})(y-y_{B,l})-\delta_{xy}\left | {\bf r} - {\bf r}_{B,l} \right |^2}{\left | {\bf r} - {\bf r}_{B,l} \right |^5}|\chi_\nu\rangle \, .
 \label{eq:v2grid}
\end{eqnarray}
\end{widetext}
The last five terms both in \ref{eq:v2grid} arise from the grid derivative. Using these definitions in eq.\ \ref{eq:qop2}, one obtains the full expression for the gradient of the charge operator including grid derivatives (see eq. S2 in the Supporting Information).

\subsection{Charge conservation}

The ESPF atomic charges have an error inherent to the finite size of the numerical grid. Usually, ESP-type methods conserve the total charge by adding a simple Lagrangian constraint in the solution of the ESP fitting equations. However, this strategy introduces extra electrostatic kernel pseudoinverses in the definition of the charges,\cite{Sigfridsson1998,Hayashi_00} which increase the complexity of equations and the computational cost of gradients and hessians.\cite{Schwinn2020} In addition, the least-square fitting of charges with Lagrangian constraint is prone to rank deficiency problems.\cite{Francl2007} To avoid this, we propose the definition of a new atomic charge operator that exactly conserves total charge without adding any extra pseudoinverses in the definition,
\begin{equation}
 {\hat Q}'_{A,\mu\nu}={\hat Q}_{A,\mu\nu}+\mathrm{N}_\mathrm{QM}^{-1}\left(S_{\mu\nu}-\sum_B^{\mathrm{N}_\mathrm{QM}}{ {\hat Q}}_{B,\mu\nu}\right) \, .
 \label{eq:qnew}
\end{equation}
As we will show in the rest of this section, this operator conserves the total charge, keeping the computational efficiency of its energy derivatives by rewriting the charge conservation in the form of an average external potential. Hereafter, we use the primed charges as indication of charge-conserved versions using the operator defined in Eq.\ \ref{eq:qnew}. In this new atomic charge operator, the ``spurious charge'' errors (i.e., the deviation of the ESPF charge from the total charge) are equivalently shared between all atoms. For large grids, this difference is close to 0, and the charge operator falls back to the original definition in Eq.\ \ref{eq:qop}. The conservation of the total charge in the new operator can be easily verified by tracing this operator with the AO density matrix, thus obtaining
\begin{equation}
 Q'_A=Q_A+\mathrm{N}_\mathrm{QM}^{-1}\left(\mathrm{N}_\mathrm{el}-\sum_B^{\mathrm{N}_\mathrm{QM}}{Q_B}\right) \, ,
\end{equation}
using the definition of eq. \ref{eq:qespf} and noting that $\mathrm{Tr}_\mathrm{AO}\left[{\bf P}{\bf S}\right]=\mathrm{N}_\mathrm{el}$, where $\mathrm{N}_\mathrm{el}$ is the total number of electrons of the QM subsystem. Therefore, the total charge conservation condition is satisfied exactly,
\begin{equation}
 \mathrm{Tr}_\mathrm{QM}\left[{\bf Q}'\right]=\mathrm{N}_\mathrm{el} \, .
\end{equation}
Using this atomic charge operator in the energy expression (eq. \ref{eq:e0espf}) leads to the new interaction energy expression
\begin{equation}
 \Delta E'=\mathrm{Tr}_\mathrm{QM}\left[{\bf q}\pmb{\phi}\right] + \Phi_\mathrm{av}\left(\mathrm{Tr}_\mathrm{QM}\left[{\bf Q}\right]-\mathrm{N}_\mathrm{el}\right) \, ,
\end{equation}
in which we defined the average external potential as $\Phi_\mathrm{av}=\mathrm{N}_\mathrm{QM}^{-1}\mathrm{Tr}_\mathrm{QM}\left[\pmb{\phi}\right]$. Similarly, the new definition of the atomic charge operator leads to additional terms in the ESPF hamiltonian (eq. \ref{eq:espfop}),
\begin{equation}
 h'_{\mu\nu}=\sum_A^{\mathrm{N}_\mathrm{QM}}{\left(\Phi_\mathrm{av}-\phi_A\right){\hat Q}_{A,\mu\nu}} -\Phi_\mathrm{av}S_{\mu\nu} \, .
\end{equation}
In principle, one could equivalently add last term as a constant $-\Phi_\mathrm{av}\mathrm{N}_\mathrm{el}$ in the energy. However, the $\Phi_\mathrm{av}\mathrm{Tr}_\mathrm{QM}[{\hat {\bf Q}}]$ alone is large and would disrupt the wavefunction. The two terms together lead to a small contribution in the interaction operator and a correct wavefunction.

\paragraph{First derivative:} 
The first derivative of the atomic charge-conserving operator is simply given by
\begin{equation}
 {\hat Q}'^{x_C}_{A,\mu\nu}={\hat Q}^{x_C}_{A,\mu\nu}+\mathrm{N}_\mathrm{QM}^{-1}\left(S^{x_C}_{\mu\nu}-\sum_B^{\mathrm{N}_\mathrm{QM}}{ Q^{x_C}_{B,\mu\nu}}\right) \, ,
\end{equation}
in which the ESPF operator charge derivative is given in eq. \ref{eq:q1op}. The atomic charge-conserving operator derivative now satisfies exactly the following condition for the full atomic charge derivative,
\begin{equation}
 \mathrm{Tr}_\mathrm{QM}\left[{\bf Q}'^{x_C}\right]=\sum_A^{\mathrm{N}_\mathrm{QM}}{\mathrm{Tr}_\mathrm{AO}\left[{\bf P}^{x_C}{\bf Q}'_A+{\bf P}{\bf Q}_A'^{x_C}\right]}=0 \, .
 \label{eq:dqsum}
\end{equation}
It is worth to note that the sum of atomic charges first derivatives at fixed MO coefficient is given by
\begin{eqnarray}
 \mathrm{Tr}_\mathrm{QM}\left[{\bf Q}'^{(x_C)}\right]&=&\sum_A^{\mathrm{N}_\mathrm{QM}}{\mathrm{Tr}_\mathrm{AO}\left[{\bf P}{\bf Q}_A'^{x_C}\right]} \nonumber \\
 &=&\mathrm{Tr}_\mathrm{AO}[{\bf P}{\bf S}^{x_C}] \, ,
 \label{eq:dqsumfix}
\end{eqnarray}
thus different from 0. For a translational invariant gradient, the following conditions should be satisfied,\cite{Dupuis1978,Kahn1981,Banarjee1985}
\begin{equation}
 \sum_C^{\mathrm{N}_\mathrm{QM}}{\mathrm{Tr}_\mathrm{QM}\left[{\bf Q}'^{(x_C)}\right]}=0 \, ,
 \label{eq:dqti1}
\end{equation}
and,
\begin{equation}
 \sum_C^{\mathrm{N}_\mathrm{QM}}{\mathrm{Tr}_\mathrm{QM}\left[{\bf Q}^{(x_C)}\right]}=0 \, .
 \label{eq:dqti2}
\end{equation}
The former is trivially satisfied, while the latter is only satisfied when grid derivatives are included (see below). 

The previous expressions for the first derivatives of the ESPF energy (eq. \ref{eq:e1espf}) and operator (eq. \ref{eq:espfop}) have additional terms due to the charge conservation. The first derivative of the energy is given by
\begin{eqnarray}
 \Delta E'^{x_C}&=&\mathrm{Tr}_\mathrm{QM}\left[{\bf Q}^{x_C}\left(\Phi_\mathrm{av}-\pmb{\phi}\right) + {\bf q}'\pmb{\phi}^{x_C}\right] \, .
\end{eqnarray}
It is important to note that the main difference between the charge-conserved gradient and the initial gradient in Eq.\ \ref{eq:e1espf} is only in the presence of the conserved charges ${\bf q}'$ and the average potential, which are straightforwardly calculated at virtually no additional computational cost. In both gradients, the charge operator first derivative is the same operator. The first derivative of the ESPF operator is given by
\begin{eqnarray}
 h'^{x_C}_{\mu\nu}&=&\sum_A^{\mathrm{N}_\mathrm{QM}}{{\hat Q}^{x_C}_{A,\mu\nu}\left(\Phi_\mathrm{av}-\phi_A\right)}-\Phi_\mathrm{av}S^{x_C}_{\mu\nu} \\
 &+&\sum_A^{\mathrm{N}_\mathrm{QM}}{{\hat Q}_{A,\mu\nu}\left(\Phi_\mathrm{av}-\phi_A\right)^{x_C}}-\Phi^{x_C}_\mathrm{av}S_{\mu\nu} \, . \nonumber
\end{eqnarray}
It is important to note that only the derivatives of the classical external potential and the AO overlap are required for charge conservation.

\paragraph{Second derivative:}
Similarly, the second derivative of the atomic charge-conserving operator is given by
\begin{equation}
 Q'^{x_Cy_D}_{A,\mu\nu}=Q^{x_Cy_D}_{A,\mu\nu}+\mathrm{N}_\mathrm{QM}^{-1}\left(S^{x_Cy_D}_{\mu\nu}-\sum_B^{\mathrm{N}_\mathrm{QM}}{Q^{x_Cy_D}_{B,\mu\nu}}\right) \, ,
\end{equation}
in which the ESPF operator charge derivative is given in eq. \ref{eq:qop2}. The atomic charge-conserving operator derivative now satisfies exactly the following condition,
\begin{equation}
 \mathrm{Tr}_\mathrm{QM}\left[{\bf Q}'^{x_Cy_D}\right]=0 \, .
\end{equation}
The second derivatives of the ESPF energy and operator with charge-conserving operators are thus given respectively by
\begin{eqnarray}
 &&\Delta E'^{x_Cy_D}=\mathrm{Tr}_\mathrm{QM}\left[{\bf Q}^{x_Cy_D}\left(\Phi_\mathrm{av}-\pmb{\phi}\right)+ {\bf q}'\pmb{\phi}^{x_Cy_D}\right] \\
 &+& \mathrm{Tr}_\mathrm{QM}\left[{\bf Q}^{x_C}\left(\Phi_\mathrm{av}-\pmb{\phi}\right)^{y_D} \right] + \mathrm{Tr}_\mathrm{QM}\left[{\bf Q}^{y_D}\left(\Phi_\mathrm{av}-\pmb{\phi}\right)^{x_C} \right] \, , \nonumber
\end{eqnarray}
and
\begin{eqnarray}
 h'^{x_Cy_D}_{\mu\nu}&=&\sum_A^{\mathrm{N}_\mathrm{QM}}{{\hat Q}^{x_Cy_D}_{A,\mu\nu}\left(\Phi_\mathrm{av}-\phi_A\right)}-\Phi_\mathrm{av}S^{x_Cy_D}_{\mu\nu} \\
 &+& \sum_A^{\mathrm{N}_\mathrm{QM}}{{\hat Q}^{y_D}_{A,\mu\nu}\left(\Phi_\mathrm{av}-\phi_A\right)^{x_C}}-\Phi^{x_C}_\mathrm{av}S^{y_D}_{\mu\nu} \nonumber \\
 &+& \sum_A^{\mathrm{N}_\mathrm{QM}}{{\hat Q}^{x_C}_{A,\mu\nu}\left(\Phi_\mathrm{av}-\phi_A\right)^{y_D}}-\Phi^{y_D}_\mathrm{av}S^{x_C}_{\mu\nu} \nonumber \\
 &+& \sum_A^{\mathrm{N}_\mathrm{QM}}{{\hat Q}_{A,\mu\nu}\left(\Phi_\mathrm{av}-\phi_A\right)^{x_Cy_D}}-\Phi^{x_Cy_D}_\mathrm{av}S_{\mu\nu} \nonumber \, .
\end{eqnarray}

\subsection{Derivatives with respect to MM-type atoms}

Up to now, we concentrated exclusively on atomic charge and energy derivatives induced by QM perturbations. Although the energy is uniquely defined for MM or QM atoms, the corresponding derivatives lead to different expressions. In Ref.\ \citenum{Schwinn2019}, we showed the specific coupled-perturbed (CP) equations for MM atom perturbations, in order to construct the density derivatives with respect to MM atoms, necessary to build up the full Hessian. A new set of CP equations, dubbed Q-vector equations, were constructed to avoid the scaling of CP with the MM size (see Eq. 20 of Ref.\ \citenum{Schwinn2019}). The same Q-vector equations are satisfied with the charge-conserving operator, in which Eq.\ \ref{eq:qnew} is used as perturbation operator instead. Hereafter, we show the energy gradient and hessian with respect to MM perturbations. In order to distinguish MM- from QM-type atom perturbations, a tilde is used for the former.

\paragraph{First derivative:}
The full gradient at the QM/MM level has two vectors, corresponding to the QM and MM derivatives.\cite{Ferre02} The QM first derivative is given by Eq.\ \ref{eq:e1espf}, while the interaction energy gradient with respect to an MM atom is simply given by
\begin{equation}
 \Delta E^{\tilde{x}_C}=\mathrm{Tr}_\mathrm{QM}\left[{\bf q}'\pmb{\phi}^{\tilde{x}_C}\right] \, .
\end{equation}
This equation stems from the fact that $Q_{A,\mu\nu}^{x_C}=0$ for any MM atom $C$, since the electrostatic kernel and the electrostatic grid are independent of the MM atom positions. In addition, the trace of the density derivative with the Fock operator, $\mathrm{Tr}_\mathrm{AO}\left[{\bf P}^{\tilde{x}_C}{\bf F}\right]$, is substituted by the equivalent term $-\mathrm{Tr}_\mathrm{AO}\left[{\bf W}{\bf S}^{\tilde{x}_C}\right]$, in which ${\bf W}$ is the energy-weighted matrix.\cite{Pople_79} Since ${\bf S}^{\tilde{x}_C}=0$ for MM atoms, this terms cancel out exactly.

\paragraph{Second derivative:}
The full Hessian at the QM/MM level has four blocks, corresponding to the QM-QM, QM-MM, MM-QM and MM-MM derivatives.\cite{Schwinn2020,Schwinn2020b} The QM-QM block is given by Eq.\ \ref{eq:e2espf}. The QM-MM and MM-QM derivatives are equivalent due to Schwarz equality of partial derivatives ($\Delta E^{\tilde{y}_Cx_D}=\Delta E^{x_C\tilde{y}_D}$). The MM-QM second derivative with charge conserving operators is thus given by
\begin{equation}
 \Delta E^{\tilde{x}_Cy_D}=\mathrm{Tr}_\mathrm{QM}\left[{\bf q}'\pmb{\phi}^{\tilde{x}_Cy_D}+{\bf Q}^{y_D}\left(\Phi_\mathrm{av}-\pmb{\phi}\right)^{\tilde{x}_C}\right] \, , 
\end{equation}
and the MM-MM second derivative by
\begin{equation}
 \Delta E^{\tilde{x}_C\tilde{y}_D}=\mathrm{Tr}_\mathrm{QM}\left[{\bf q}'\pmb{\phi}^{\tilde{x}_C\tilde{y}_D}+{\bf Q}^{\tilde{y}_D}\left(\Phi_\mathrm{av}-\pmb{\phi}\right)^{\tilde{x}_C}\right] \, .
\end{equation}

\section{Computational details}

The new charge conserving and grid derivative including ESPF model is capable of calculating the energy, its gradient and hessian with respect to all, QM and MM, degrees of freedom. This ESPF v2.0 is currently implemented in a local development version of {\sc Gaussian}16,\cite{g16} interfaced with a modified version of Tinker 8.7.1.\cite{tinker} Unless otherwise stated, all calculations presented below are performed at the B3LYP/6-31G* and Amber99 levels of theory for QM and MM respectively.\cite{b3,lyp,vwn,g631,amber18} The implementation has been tested by comparing numerical and analytic, which lead to an average absolute differences between $10^{-3}-10^{-5}$ a.u. in all cases.

The atom-centered grids have been constructed from Lebedev spheres, in which all weight has been set to 1 and the weight derivatives to 0. Each atom features three spheres of 100 Lebedev points each, that have been equally spaced using the van der Waals radius of the atom. Each grid point inside the van der Waals radius of any of the atom of the molecule is excluded. No scaling of the QM-MM electrostatic interactions was applied, while we maintained the standard scaling factors for the purely MM electrostatic interactions.

The construction of the ESPF operator elements and its derivatives requires the computation of on-grid electrostatic integrals up to the second order in the interaction kernel, as well as integral derivatives up to the first order [see eqs.\ \ref{eq:v1grid} and \ref{eq:v2grid}], which are already available in {\sc Gaussian}16. The derivatives of the electrostatic kernel matrix $\bf{T}$ are formed on the fly and contracted with the integrals and the pseudoinverse matrices. The pseudoinverse ${\bf T}_w^+$ has been computed by means of a Cholesky decomposition of $({\bf T}^\dagger{\bf w}{\bf T})$ before its inversion. The intermediate matrix $B_{B,l}=\sum^{\mathrm{N}_\mathrm{QM}}_A{\phi_A \left[T_w^+\right]_{A;B,l}}$ is stored in memory and then contracted with the integrals with a sum over grid points, leading directly to an energy contribution.

\section{Results and Discussion}

\subsection{Grid effect on the atomic charges}

\begin{figure}
 \includegraphics[width=0.7\linewidth,keepaspectratio]{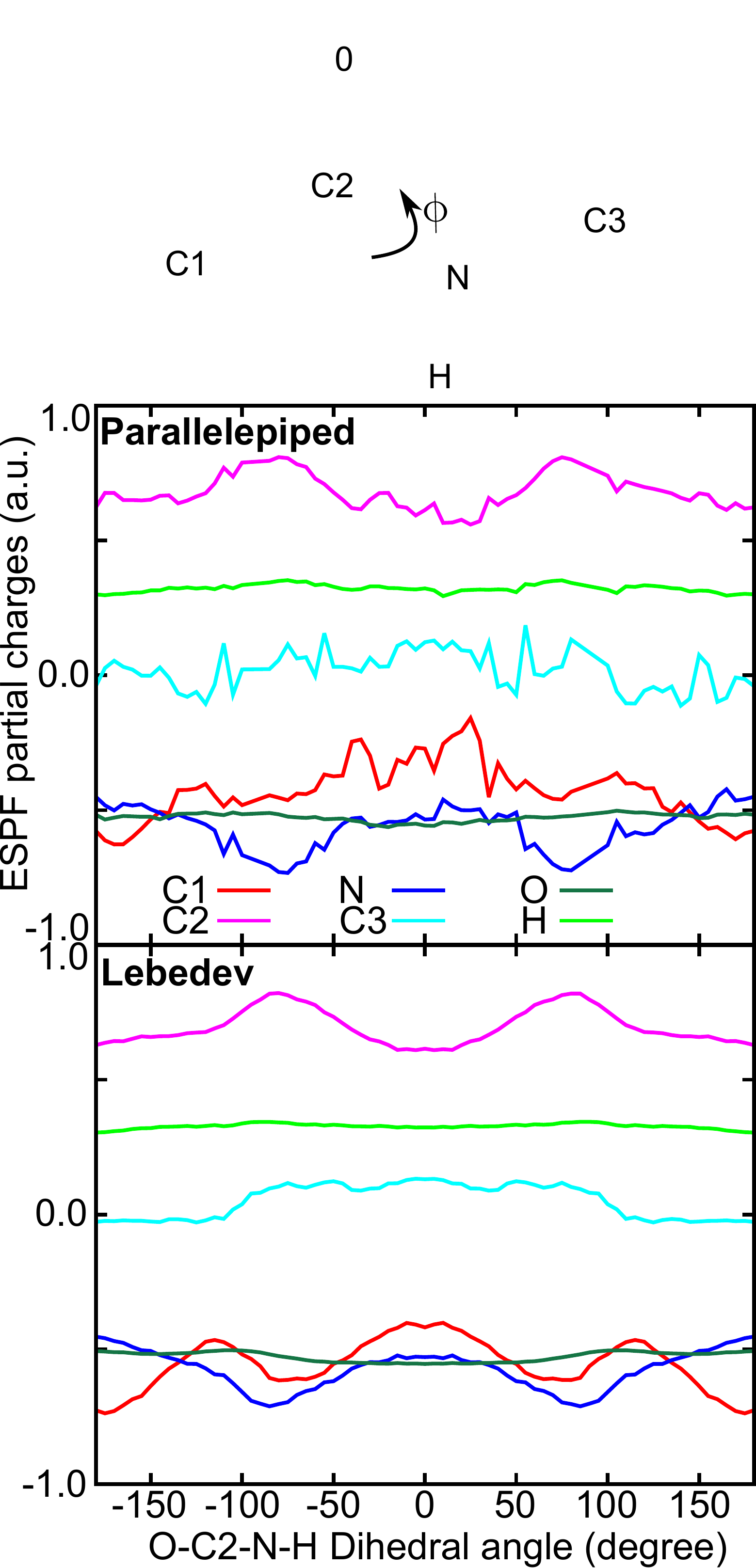}
 \caption{ESPF partial atomic charges (in a.u.) along the O--C2--N--H dihedral angle of N-methylacetamide using a parallelepiped grid (1777 grid points) or a Lebedev grid (2572 grid points) using charge conserving operators. For simplicity, the charges of the methyl hydrogen atoms have been omitted.}
 \label{fig1}
\end{figure}

The grid type selected to derive atomic charges fitted to the electrostatic potential largely impacts their values and variations when the molecular geometry is modified.\cite{Breneman1990} This was clearly shown by \citeauthor{Hu2007} in Ref. \citenum{Hu2007}, who computed atomic charges along one dihedral angle in N-methylacetamide. In Figure \ref{fig1}, we take the same example and apply it to ESPF charges with (ESPF v1.0) parallelepiped and (ESPF v2.0) Lebedev grids.
In this case, no MM atom is present in the calculation. As it can be clearly seen in Figure \ref{fig1}, the numerical values of each atomic charge remain close whatever the grid type. However, their variations with the $\phi=$O--C2--N--H dihedral angle are clearly different. Using the Lebedev grid leads to a rather continuous and smooth change of the charges while the parallelepiped grid introduces an appreciable numerical noise. In addition, the charge variation should be symmetric with respect to $\phi=0^{\circ}$, i.e. $q_A(\phi)=q_A(-\phi)$ for any atom $A$. This property is perfectly satisfied by the Lebedev grid but not by the parallelepiped grid. The reason for these differences is essentially rooted into the symmetry of the numerical grid. Since a Lebedev grid is constructed from spheres centered on nuclei, atomic displacements distort evenly the grid, hence keeping a consistent overall molecular symmetry. Conversely, a parallelepiped grid is defined using three vectors from the most extreme atom Cartesian coordinates in the three space directions. If the atoms at the extremes of the molecule are kept fixed while other move, the orientation of the grid does not change, thus introducing a numerical noise in the computation of charges (or any other grid-dependent property).

\subsection{Grid effect on the atomic charge derivatives}

\begin{figure}
 \includegraphics[width=0.54\linewidth,keepaspectratio]{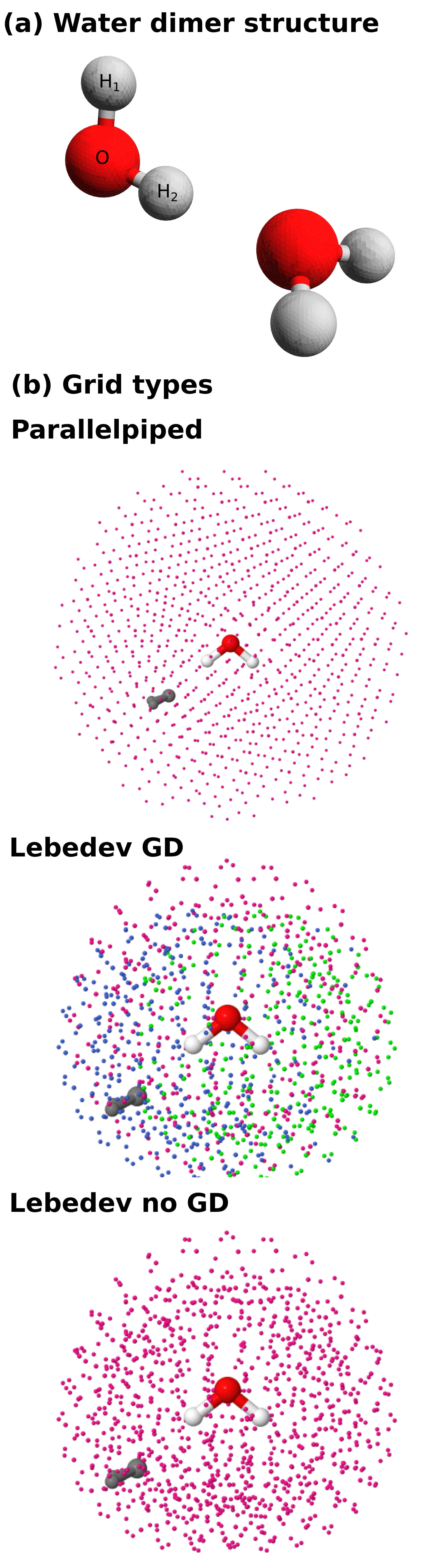}
 \caption{(a) (H$_2$O)$_2$ model used for the QM/MM calculations. The QM atoms are numbered. (b) Parallelpiped and Lebedev grids used to expand the electrostatic integrals. The QM and MM water molecules are shown respectively in colored and grey ball and sticks. For Lebedev grid with grid derivatives (GD), the grid points are colored in magenta, green and blue for grid points centred on O1, H1 and H2 respectively. When all points are in magenta, no atom origin is specified.}
 \label{fig2}
\end{figure}

The consideration of the grid derivatives are important in order to satisfy certain sum rules of the atomic charge derivatives. The sum rule defined in eq.\ \ref{eq:dqsum} ensures that the total charge of the QM subsystem is kept constant and no electron density spilling effect occurs between the QM and MM subsystems. In Table \ref{tab:sumqx}, we computed this sum rule for a QM/MM water dimer ( Figure \ref{fig2}) using either a parallelepiped grid, hence excluding the grid derivatives, or a Lebedev grid, hence including or excluding the grid derivative terms.

Only the selection of Lebedev grids, including the grid derivative contributions, is able to retain the translational invariance of the QM/MM Hamiltonian. When the grid derivatives are excluded, the corresponding charge and gradient sum rules are only satisfied approximately. The magnitude of the spurious extra charge is independent of the grid size. Indeed, these sum rule deviations are still observed when using a finer grid, with a similar magnitude as for the coarser grids. This shows that the grid derivatives are essential for satisfying the derivative sum rules. If one compares the grid-derivative formulas (eqs. \ref{eq:t1grid} and \ref{eq:v1grid}) with the formulas without these contributions (eqs. \ref{eq:t1} and \ref{eq:v1}), a physical interpretation can be given. Grid derivative terms introduce dipolar terms into the electrostatic integral derivatives, resulting in extra dipole-charge type interactions between the perturbed atom and the rest of atomic charges of the QM subsystem.

\begin{table}
 \centering (a) Charge derivative conservation
 \begin{tabular}{lcccc}
 \hline
Grid	&	\# Points	&	x	&	y	&	z	\\
\hline
Paral. (no GD)$^1$	&	885	&	-0.180	&	-0.321	& \-	0.014	\\
 		&	56385	&	\- 0.012	& \- 0.071		&\- 0.120	\\
	\hline
Lebedev (GD)$^2$&	928	&\-	0.000	&\-	0.000	&\-	0.000	\\
 		&	58895	&\-	0.000	&\-	0.000	&\-	0.000	\\
\hline
Lebedev	(no GD)$^1$&	928	&\-	0.057 	&	-0.091	&\-	0.354	\\
 	&	58895	&	\- 0.133	&	-0.001	&\- 0.314		\\
\hline
 \end{tabular}\newline
 
 \centering (b) Gradient translational invariance
 \begin{tabular}{lcccc}
 \hline
Grid	&	\# Points	&	x	&	y	&	z	\\
\hline
Paral.	(no GD)$^1$ &	885	&	\- 18.560	&	-16.289	& \-	20.693	\\
 		&	56385	&	\- 14.466	& -7.813		&\- 22.430	\\
	\hline
Lebedev (GD)$^2$ &	928	&\-	0.012	&\-	0.014	&\-	0.001	\\
		&	58895	&\-	0.012	&\-	0.014	&\-	0.001	\\
\hline
Lebedev	 (no GD)$^1$&	928	&\-	9.535 	&	-9.859	& \- 7.111	\\
	&	58895	&	\- 10.067	&	-8.642	&\- 7.334		\\
\hline
 \end{tabular}\newline
 $^1$ No grid derivative included
 $^2$ Grid derivative included
 \caption{(a) Deviation from the ideal sum rule of the charge derivatives (Eq.\ \ref{eq:dqti2}) for a QM/MM water dimer model (Figure \ref{fig2}) using different numerical grids including or not the grid derivative (GD) contributions. Units are 10$^{-3}$ charge a.u./Bohr. (b) Sum of each atom QM/MM energy gradient contribution along one space direction. When the translational invariance is satisfied, each sum should be 0 (eq. \ref{eq:tinv}). Units are meV/\AA{}.}
 \label{tab:sumqx}
\end{table}

A consequence of not satisfying the charge derivative sum rule is the breaking of the QM/MM Hamiltonian translational invariance. When this condition is satisfied, it translates to
\begin{equation}
 \sum_C^{\mathrm{N}}{E^{x_C}}=0 \, .
 \label{eq:tinv}
\end{equation}
The sum over $N$ runs over all atoms. Table \ref{tab:sumqx}(b) reports values for the gradient translational invariance for the different numerical grids. One could think that the use of atom-centred Lebedev grids, which are invariant under translations by construction, suffice to guarantee the exact translational invariance of the QM/MM Hamiltonian. As it can be observed, this is not the case, irrespective of the grid size. A similar trend is observed for parallelpiped grids. Only Lebedev grigds including grid derivative terms lead to a clear translational invariance of the QM/MM hamiltonian.

As an example of the effect of the grid derivatives, we compare in Table\ \ref{tab:freq} the 12 frequencies characterizing the same water dimer (Figure \ref{fig2}) normal modes obtained with a full QM calculation with those obtained using QM/MM.

\begin{table}
 \centering
 \begin{tabular}{ccccc}
 \hline
Mode	&	QM	&	Leb. 	&	Leb. 	&	Paral. 	\\
	&		&	GD	&	no GD	&	no GD	\\
\hline
1 & -0.04 & -30.43 & -41.68 & -23.94 \\
2 & 0.01 & -20.94 & -21.41 & -16.30 \\
3 & 0.03 & -6.98 & -12.45 & -11.71 \\
4 & 2.55 & 4.22 & -4.66 & -7.65 \\
5 & 2.63 & 20.09 & -0.02 & 10.94 \\
6 & 17.89 & 30.01 & 10.63 & 29.17 \\
\hline
7 & 59.26 & 68.78 & 49.32 & 69.20 \\
8 & 165.54 & 160.80 & 160.23 & 160.92 \\
9 & 180.32 & 202.77 & 201.64 & 199.75 \\
10 & 218.66 & 228.55 & 225.29 & 227.12 \\
11 & 434.46 & 412.38 & 402.99 & 404.42 \\
12 & 659.11 & 656.41 & 653.15 & 655.76 \\
13 & 1671.71 & 1708.27 & 1712.72 & 1711.37 \\
14 & 1705.11 & 2340.07 & 2340.10 & 2340.08 \\
15 & 3698.50 & 3659.94 & 3663.02 & 3660.91 \\
16 & 3801.66 & 3706.01 & 3707.46 & 3706.44 \\
17 & 3880.08 & 3739.61 & 3739.55 & 3739.58 \\
18 & 3903.38 & 3872.34 & 3871.81 & 3871.26 \\
\hline
 \end{tabular}
 \caption{Comparison of the full spectrum of harmonic frequencies for the water dimer ( Figure\ \ref{fig2}) at the minimum energy structure of the full QM model and the QM/MM ESPF models using different grids. The Cartesian coordinates of each minimum energy structure can be found in supporting information. The first 6 vibrations correspond to translations and rotations and should be close to 0. Frequencies have been computed at the B3LYP/6-31G(2d,2p) level. The frequencies are given in cm$^{-1}$.}
 \label{tab:freq}
\end{table}

Whatever the grid type, most of the QM/MM frequencies are very close to each other. Moreover, they are similar to (but not perfectly matching) the QM spectrum. The first 6 vibrations correspond to translations and rotations, and should be close to 0. In this case, the inclusion of grid derivatives brings the translations close to the 0 frequency, better than when grid derivatives are not included. In all QM/MM cases, rotations are not perfectly separated, since this would require the implementatino of a rotational invariant Lebedev grid.\cite{Johnson1994} For the rest of frequencies, the difference with the QM calculation can be attributed to the quantitatively different force constants in the selected classical force field. However, note that recent developments in polarizable force fields show a better comparison with respect to quantum calculations of force constants.\cite{Semrouni2014,Spicher20}

\subsection{Reoxidation of fully reduced anionic flavin adenine dinucleotide in cryptochrome}

\begin{figure}
 \includegraphics[width=\linewidth,keepaspectratio]{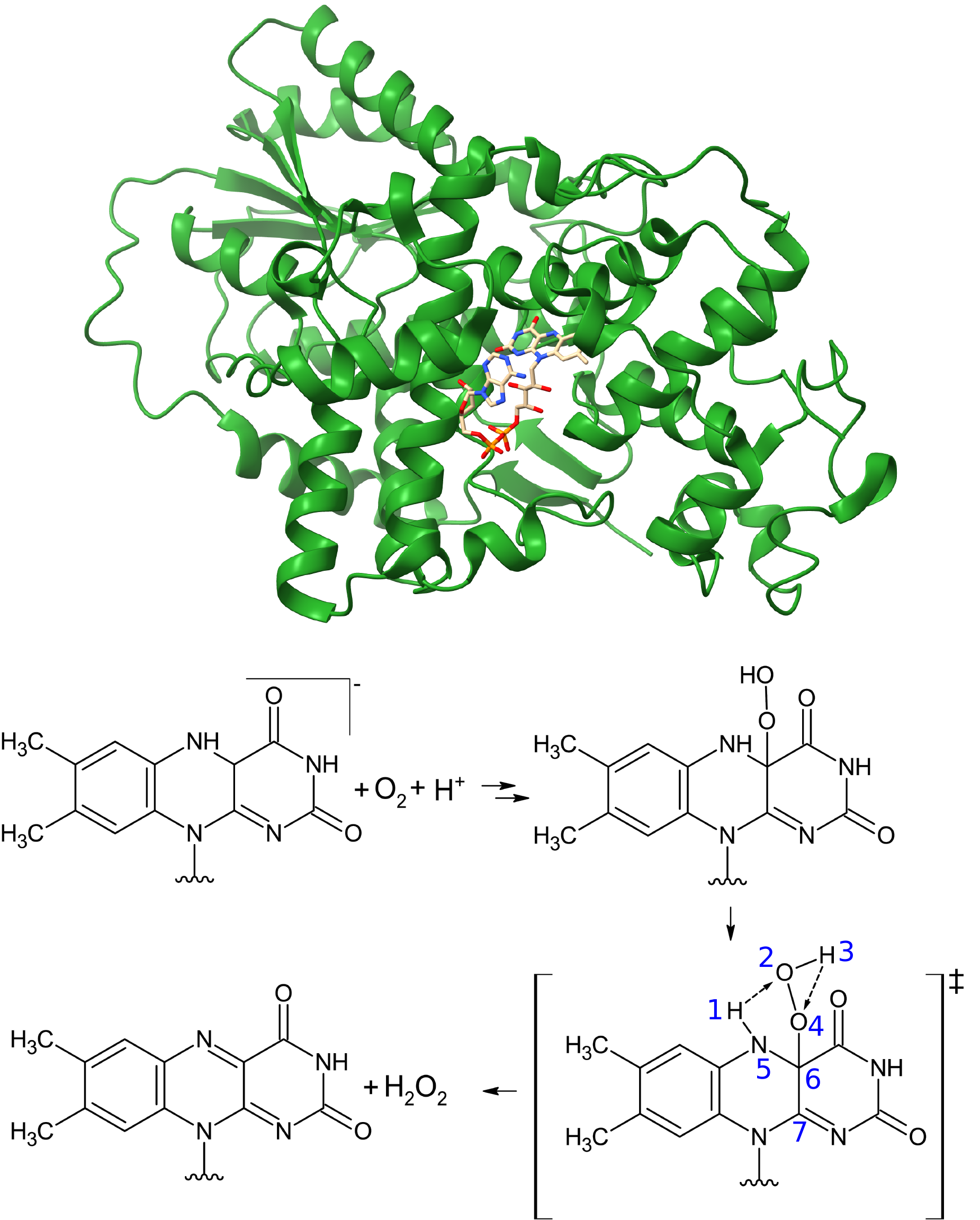}
 \caption{(top) Structure of \textit{Arabidopsis thaliana} cryptochrome 3 used in the QM/MM model (PDBID: 2J4D). The QM subsystem consists of flavin adenine dinucleotide (in ball and stick), while the rest of the protein has been treated at the MM level. The reduced form of FAD and the hydroperoxyl molecules have been placed in a second step, also treated at the QM level. (bottom) Reoxidation reaction considered in this study. The important atoms are numbered in blue on the representative structure of the transition state. The structures of the transition states can be found in Supporting Information.}
 \label{fig3}
\end{figure}

Cryptochromes (see Figure\ \ref{fig3}) are flavoproteins acting as blue light receptors, potentially at the origin of magnetoreception,\cite{Schulten1978} i.e., the sensing of the geomagnetic field. This mechanism could be possibly happening in the protein cavity upon blue light absorption, by generating a biradical between a tryptophan triad and the flavin adenine dinucleotide (FAD).\cite{Solovyov2012} Alternatively, it has been proposed that biradicals can be formed during the reoxidation reaction of fully reduced flavin adendine dinucleotide (FADH$^-$) with molecular oxygen.\cite{Mueller2011} Recently, one of us showed that such radicals can be stabilized in the protein FAD pocket for a long time, due to strong electrostatic interactions.\cite{Mondal2019} The complete reaction is complex, since several electron transfers can occur between reduced FAD forms and oxygen when the complex is activated with light.\cite{Mondal2019} In the hypothesis of a mechanism featuring reactive oxygen species, the biradical should be stable for at least several milliseconds for the magnetoreception to be effective. Here, we study the final step of the oxidation reaction of reduced FAD by molecular oxygen in a cryptochrome protein, namely, the formation of hydrogen peroxide. This reaction step is irreversible, and thus, a rapid formation of \ce{H2O2} could contradict the possibility of oxygen-flavin-based magnetoreception. 

The reaction mechanism considered in this study is depicted in Figure\ \ref{fig3}. After having optimized the geometry of the reactant, we have been able to locate the corresponding transition state, which corresponds to the simultaneous proton transfer from the secondary amine to one \ce{-OOH} oxygen, and from the same oxygen atom to the other one in \ce{-OOH}. The QM/MM imaginary frequency of this mode is 1476i $cm^{-1}$ in the protein. The activation free energy at standard conditions of temperature and pressure is 48.58 kcal$\cdot$mol$^{-1}$, leading to a reaction rate of 1.59$\cdot$10$^{-23}$ s$^{-1}$. In the gas phase, a similar transition state exists for lumiflavin (7,8,10-trimethylisoalloxazine) with an imaginary frequency of 1356i $cm^{-1}$ and a free energy of activation of 8.78 kcal$\cdot$mol$^{-1}$, leading to a reaction rate of ~2.27$\cdot$10$^{6} $ s$^{-1}$. Indeed, this show that this reoxidation reaction is forbidden, while in gas phase is fast. This shows that reoxidation reaction between oxygen and FAD should occur via a different mechanism in protein (see Ref.\ \citenum{Mondal2019}). In Table\ \ref{tab:tsdist}, the geometries of the two transition states for lumiflavin in gas phase and FAD in cryptochrome are compared (for the Cartesian coordinates and the ESPF charges, see Supporting Information). As it can be seen, most geometrical parameters are similar. The O-O distance is almost the same in both cases, close to the equilibrium value of 1.45 \AA{} for H$_2$O$_2$ (computed at the MP2/cc-pVQZ level in the gas phase). In protein, the oxygens are further apart from FAD ($\sim$0.5 \AA{} further from carbon and $\sim$0.4 \AA{} further from nitrogen) with respect to gas phase. The increase of the $C6-O4$ distance destabilizes the transition state complex, and can be attributed to a redistribution of charge. The charge on O4 net changes by $\sim$-0.12 a.u. of charge between the minimum and the transition state, C6 by $\sim$-0.4 a.u., while C7 is $\sim$+0.14. Other atoms in the ribitol moiety also have a net positive charge increase (see Supporting Information). In addition, the hydrogens are closer to the equilibrium value of 0.97 \AA{} for H$_2$O$_2$ (computed at the MP2/cc-pVQZ level in the gas phase) than the gas phase transition state.

\begin{table}
 \centering
 \begin{tabular}{ccc}
 \hline
Parameter	&	Lumiflavin	&	Cryptochrome \\
\hline
O2-O4 & 1.522 & 1.588 \\
O4-C6 & 1.759 & 2.484 \\
C6-C7 & 1.501 & 1.454 \\
H1-N5 & 1.220 & 1.657 \\
H1-O2 & 1.289 & 1.019 \\
H3-O2 & 1.302 & 1.086 \\
H3-O4 & 1.311 & 1.147 \\
N5-H1-O2 & 76.537 & 82.387 \\
\hline
 \end{tabular}
 \caption{Comparison of main geometrical parameters of the transition states at the QM/MM level for cryptochrome and at the QM level for lumiflavin. Distances are in \AA{} and angles in degrees. Atom numbering is shown in Figure \ref{fig3}. The structures of the transition states can be found in Supporting Information.}
 \label{tab:tsdist}
\end{table}

\subsection{Computational scaling: chignolin in water droplets}

\begin{figure}
 \includegraphics[width=\linewidth,keepaspectratio]{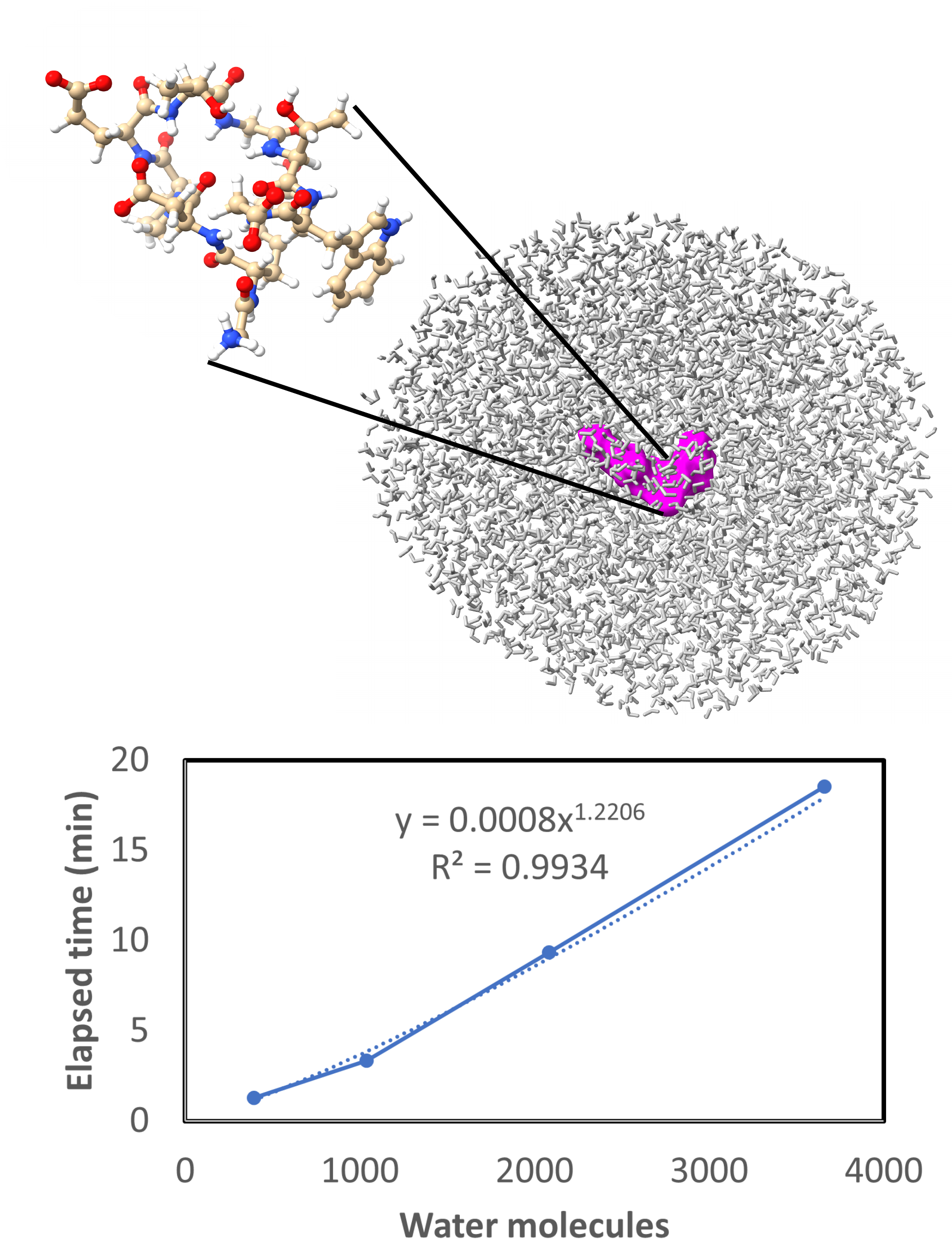}
 \caption{Scaling of the QM/MM Hessian calculations with respect to the MM atom size. (top) QM/MM model used for the scaling calculations. The QM subsystem is formed by chignolin (PDBID: 1UAO, depicted in magenta soaked in water and its structure in the offset) treated at the HF/3-21G* level and spheres of water molecules around it are computed at the MM level using Amber99. (bottom) Plot of the elapsed time (in minutes) with respect to the number of MM atoms. From these, we have substracted the elapsed time of the pure QM hessian calculation, 15.5 min. Calculations have been performed on 64 CPUs Intel(R) Xeon(R) Platinum 8280 CPU @ 2.70GHz and 2800 Gb of RAM memory.}
 \label{fig4}
\end{figure}

In Figure \ref{fig4}, we report the computational scaling of the QM/MM Hessian with respect to the number of MM atoms. As a test case, we have taken chignolin, a 10 aminoacid protein forming a double beta sheet. This molecule, consisting of 138 atoms treated at the QM level of theory. has been soaked in different microscopic water droplets with a radius of 10, 15, 20 and 25 \AA{} around any of chignolin atoms. For the sake of safe comparisons, the complete geometries have been kept frozen. As it can be seen from the trend line drawn in the graph of Figure \ref{fig4}, the scaling is $\mathcal{O}^{1.22}$, hence close to linear scaling. However, note that Figure \ref{fig4} reports timings for the building of all the QM/MM Hessian contributions. In other words, the I/O timings are not considered. Indeed, while the MM Hessian has a negligible computational cost with respect to the QM calculation, in our current implementation, Tinker transfers huge amounts of information to Gaussian16 by means of formatted text files. These I/O transfers increase the overall computational time to $\mathcal{O}^{1.85}$, ie. almost proportional to the size of the Hessian. This is however a technical problem, that could be solved by making the MM and QM programs share the same physical memory thanks to a suitable Application Programming Interface or by means of binary files. Finally, one should note that this scaling is possible thanks to the Q-vector method, which avoids the scaling of the coupled perturbed equations with the MM subsystem size.\cite{Schwinn2019,Schwinn2020,Schwinn2020b}

\section{Conclusion}

We have refined a QM/MM method based on new Electrostatic Potential Fitted (ESPF) atomic charge operators that are built to conserve the total molecular charge. This ESPF v2.0 approach is capable of calculating QM/MM energy, gradient and full Hessian. When interacting with any kind of external electrostatic potential, the ESPF operators can be added as expectation values to the total energy in a mechanical embedding approach, or directly included into the Fock operator of any self-consistent field procedure in an electrostatic embedding approach. 

The QM atomic charges are fitted to reproduce electrostatic integrals computed on a grid surrounding the QM atoms. We showed that, unlike other grids, atom-centred grids are superior to produce correct atomic charge and energy derivatives. Indeed, they avoid the numerical instabilities which are frequently appearing with other grid types when the QM geometry is distorted. In addition, the atomic charge derivative sum rules as well as the translational invariance condition are retrieved when atom-centred grids are used. Still, rotational invariance of the QM/MM Hamiltonian is not completely satisfied. This would require, for example, the use of rotational invariant Lebedev's grid.\cite{Johnson1994} However, this would require the implementation of new types of integrals, beyond the scope of the present work.

While some technical improvements are still possible for achieving a better computational efficiency (notably in terms of I/O), we have been able to locate the transition state characterizing a complex two proton transfer reaction occurring in a large cryptochrome protein.

The energy and analytic derivatives of the ESPF method with grid derivatives and charge conservation only affect the QM atomic charges, and thus they were derived for a general MM force field. In our applications, we considered only non-polarizable force fields (fixed point charge representation). The extension to polarizable force fields will change the analytic form for the external potential $\phi$ as well as its derivatives, depending on how polarization is introduced in the classical force field (Drude particles, multipolar expansion, etc.), but does not affect the ESPF expressions as long as the external MM field is independent of the QM charges.

To conclude, we formulated a general QM/MM method based on ESPF charges without approximations besides the truncation of the multipolar expansion. All the working equations shown here for the lowest order approximation (i.e., atomic charges) can be straightforwardly derived for higher-order terms.

\begin{acknowledgments}
The authors acknowledge financial support by the Agence Nationale pour la Recherche through the project BIOMAGNET (ANR-16CE29-0008-01). Centre de Calcul Intensif d'Aix-Marseille is acknowledged for granting access to its high performance computing resources.
\end{acknowledgments}

\section*{Additional information}
Supporting information includes the full expressions for the first and second derivatives of the charge operator with grid derivatives, the cartesian coordinates of all optimized structures, the ESPF charges for the minimum structure and the transition state in protein, as well as the comparison between numerical and analytic gradient and hessian for all grid types. This information is available free of charge via the Internet at \url{http://pubs.acs.org}.

\bibliographystyle{achemso}
\bibliography{refs}

\end{document}